# Correlations in 2D electron gas at arbitrary temperature and spin polarizations

Nguyen Quoc Khanh

*Department of Theoretical Physics, National University in Ho Chi Minh City, 227-Nguyen Van Cu Str., 5th District, Ho Chi Minh City, Vietnam*

___

**Abstract**

Using the Singwi-Tosi-Land-Sjolander theory we have studied the many-body effects in the two-dimensional electron gas with arbitrary polarization at finite temperatures. We have calculated the structure factors, pair correlation functions, local-field factors and the Helmholtz free energy for different values of spin polarization, temperature and density parameter. We have shown that the spin polarization and finite temperature effects are remarkable and in the low temperature or paramagnetic case our results match closely with those obtained in earlier papers .

___

## 1. Introduction

In the course of the last decades, many theoretical works have been devoted to quasi-two-dimensional systems such as electrons in inversion layers, heterojunctions and quantum wells [1]. Most of the theoretical calculations have been performed in the framework of the random phase approximation (RPA) which gives good results only in the high-density limit. In addition, it has been shown that the many-body correlations are more pronounced in two than in three dimensions [2]. The beyond-RPA effects in bulk and low-dimensional systems have been studied by several authors through different approaches[3-5]. However, most of these works referred to the paramagnetic state only and the extension to spin-polarized systems is motivated not only by fundamental many-body considerations, but also by practical experimental needs [6]. In fact the number of theoretical investigations on spin-polarized systems is growing in the recent years [7-9]. Davoudi and Tosi (DT) have studied the spin polarization in a two-dimensional electron gas (2DEG) at zero temperature [10] using the self-consistent approximation of Singwi, Tosi, Land and Sjolander ( STLS ). The finite temperature effects have been studied by Shweng and Bohm (SB) using the STLS scheme for unpolarized 2DEG [11]. It was shown that the properties of 2DEG depend remarkably on the temperature. In this paper we extend the DT's work to the 2DEG at finite temperatures. Our model is the 2DEG interacting with the $e^2/r$ law and moving in a plane over a uniform neutralizing background of positive charge. Confirming that our results reduce to those of zero temperature [10] or unpolarized case [11], we analyze the correlation characteristics of 2DEG for different values of spin-polarization, temperature and density.

## 2. Theory

We consider a 2DEG at temperature $T$ containing two spin species of surface densities $n_\uparrow$ and $n_\downarrow$ ( $n_\uparrow \geq n_\downarrow$ ) with the total density $n = n_\uparrow + n_\downarrow$ and spin polarization $\zeta = (n_\uparrow - n_\downarrow)/n$ . We will use the effective atomic units with the effective Bohr radius $a_B^* = \dfrac{\varepsilon_o \hbar^2}{m^* e^2}$ where $\varepsilon_o$ is the background dielectric constant. The STLS self-consistent equations for spin-polarized 2DEG read [10-11]

$$\chi_{\sigma\sigma'}(k,\omega) = \frac{\chi_{0\sigma}(k,\omega)[\delta_{\sigma\sigma'} + (-1)^{\delta_{\sigma\sigma'}} \psi_{\overline{\sigma}\overline{\sigma'}}(k)\chi_{0\overline{\sigma'}}(k,\omega)]}{\Delta(k,\omega)} \qquad (1)$$

$$G_{\sigma\sigma'}(k) = -\frac{1}{\sqrt{n_\sigma n_{\sigma'}}} \int \frac{d^2\vec{q}}{(2\pi)^2} \frac{\vec{k}.\vec{q}}{k.q} \left[ S_{\sigma\sigma'}\left(|\vec{k}-\vec{q}|\right) - \delta_{\sigma\sigma'} \right] \quad (2)$$

$$S_{\sigma\sigma'}(q) = -\frac{\hbar}{\pi\sqrt{n_\sigma n_{\sigma'}}} \int_0^\infty \mathrm{Im}[\chi_{\sigma\sigma'}(q,\omega)] \coth(\frac{\hbar\omega}{2k_B T}) d\omega \quad (3)$$

where

$$\Delta(k,\omega) = [1 - \psi_{\uparrow\uparrow}(k)\chi_{0\uparrow}(k,\omega)][1 - \psi_{\downarrow\downarrow}(k)\chi_{0\downarrow}(k,\omega)] - \psi_{\uparrow\downarrow}(k)\psi_{\downarrow\uparrow}(k)\chi_{0\uparrow}(k,\omega)\chi_{0\downarrow}(k,\omega) \quad (4)$$

$$\psi_{\sigma\sigma'}(k) = V_k[1 - G_{\sigma\sigma'}(k,\omega)] . \quad (5)$$

with $V_k$ is the bare electron-electron interaction , $\chi_{0\sigma}(k,\omega)$ are the susceptibilities of the ideal 2DEG and $\bar{\sigma}$ denotes the spin orientation opposite to $\sigma$.

The pair distribution function ( PDF) can be calculated from the structure factors as

$$g_{\sigma\sigma'}(r) = 1 + \frac{1}{\sqrt{n_\sigma n_{\sigma'}}} \int_0^\infty \frac{d^2 q}{(2\pi)^2} [S_{\sigma\sigma'}(q) - \delta_{\sigma\sigma'}] e^{i\vec{q}\vec{r}} \quad (6)$$

The exchange-correlation and total Helmholtz free energy are given by

$$\frac{F_{xc}}{n} = \pi n \int_0^1 d\lambda \int dr [g(\lambda,r) - 1] \quad (7)$$

$$F_{tot} = F_0 + F_{xc} \quad (8)$$

where $r_s$, $\lambda$ and $g(\lambda, r)$ are the density parameter, coupling constant and spin-averaged PDF, respectively [8-9] . Here $F_0 = F_0^\uparrow + F_0^\downarrow$ is the total free energy of a non-interacting electron gas and

$$F_0^\sigma = n_\sigma \mu_0^\sigma - E_0^\sigma \quad (9)$$

$$\mu_0^\sigma = T \log[e^{E_F^\sigma/T} - 1] \quad (10)$$

$$E_F^\sigma = 2\pi n_\sigma = (1 \pm \zeta)\pi n \quad (11)$$

$$E_0^\sigma = \frac{T^2}{\pi} \int_0^\infty \frac{dx\, x}{\exp[x - \mu_0^\sigma/T] + 1} \quad (12)$$

$$E_0 = E_0^\uparrow + E_0^\downarrow \quad (13)$$

At $T = 0$ we have

$$\frac{E_0}{n} = \frac{1+\zeta^2}{r_s^2} \tag{14}$$

We have solved the STLS self-consistent equations numerically and the results will be discussed in the next section.

## 3. Results and discussions :

In this section we report our numerical results for the structural functions and the free energy at different values of spin-polarization $\zeta$, temperature $T$ and density parameter $r_s = \dfrac{1}{a_B^* \sqrt{\pi n}}$.

*3.1. Static structure factor :*

In Figs. 1 and 2 we show the parallel and antiparallel spin structure factors $S_{\sigma\sigma'}(q)$ as function of $q/q_F$, first on varying $\zeta, T$ at $r_s = 2$ and then on varying $\zeta, r_s$ at $T = T_F$. We observe remarkable influence of spin-polarization $\zeta$, temperature $T$ and density parameter $r_s$ on the structure factors.

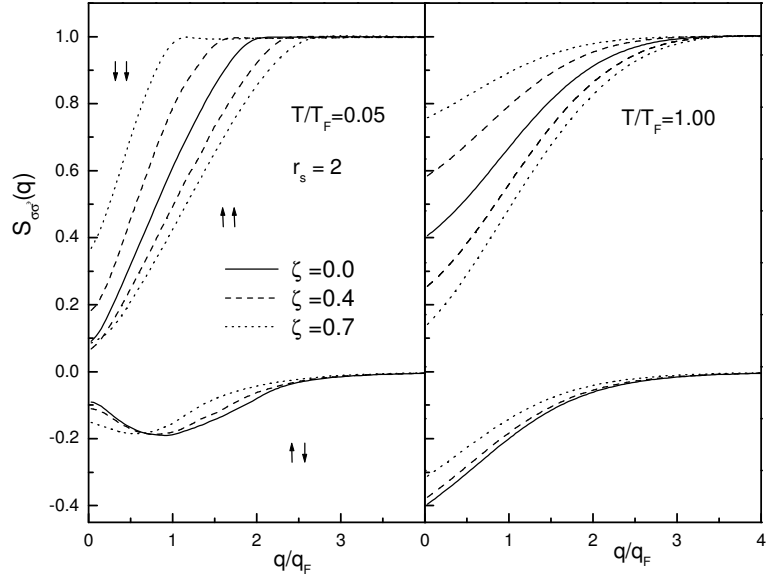

Fig. 1. Parallel and antiparallel spin structure factors $S_{\sigma\sigma'}(q)$ versus $q/q_F$ for $T = 0.05T_F$ ( left) and $T = T_F$ ( right ) and various $\zeta$ at $r_s = 2$.

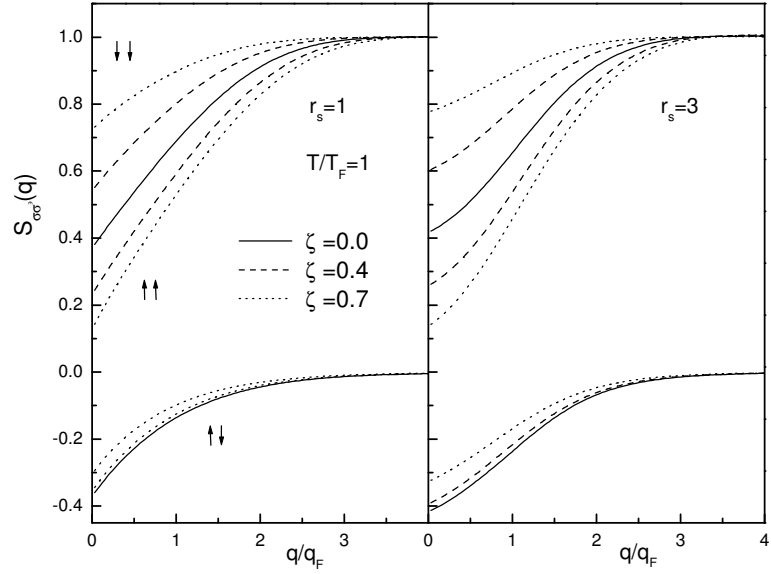

Fig. 2. Parallel and antiparallel spin structure factors $S_{\sigma\sigma'}(q)$ versus $q/q_F$ for $r_s = 1$ (left) and $r_s = 3$ (right) and various $\zeta$ at $T = T_F$.

The Fig. 3 shows the temperature and spin-polarization dependence of the partial spin structure factors $S_{\sigma\sigma'}(q)$ at $r_s = 1$. It is seen that the temperature effect is considerable and in the paramagnetic case our results are almost identical to those of SB [11]. Our SSFs for $r_s = 2$ at low temperature $T = 0.05T_F$ show similar behaviour as it is given in the DT's work for $T = 0$.

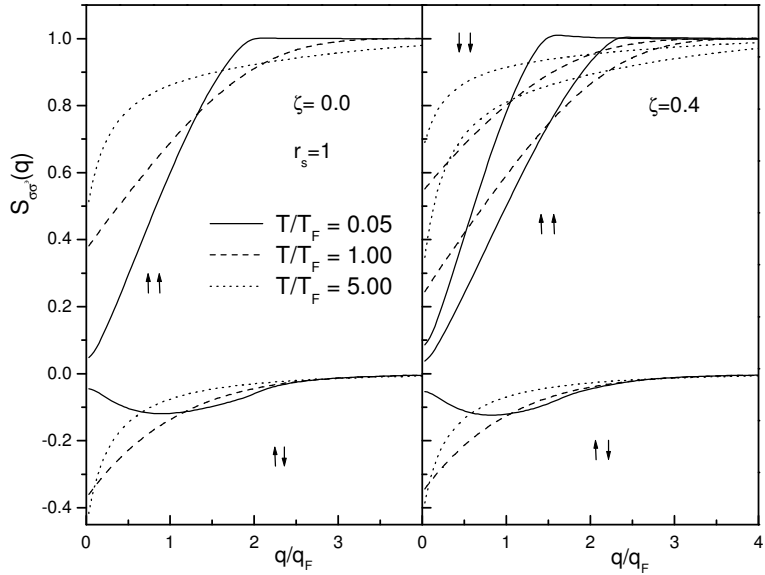

Fig. 3. Parallel and antiparallel spin structure factors $S_{\sigma\sigma'}(q)$ versus $q/q_F$ for $\zeta = 0$ (left) and $\zeta = 0.4$ (right) and various $T$ at $r_s = 1$.

The spin-averaged structure factor $S(q)$ for $r_s = 1$ and $r_s = 2$ and various $T$ at $\zeta = 0.4$ is plotted in Fg. 4 which shows the strong temperature dependence of SSF at high $T$. This behavior of SSF is similar to that shown in the SB's work for the unpolarized 2DEG.

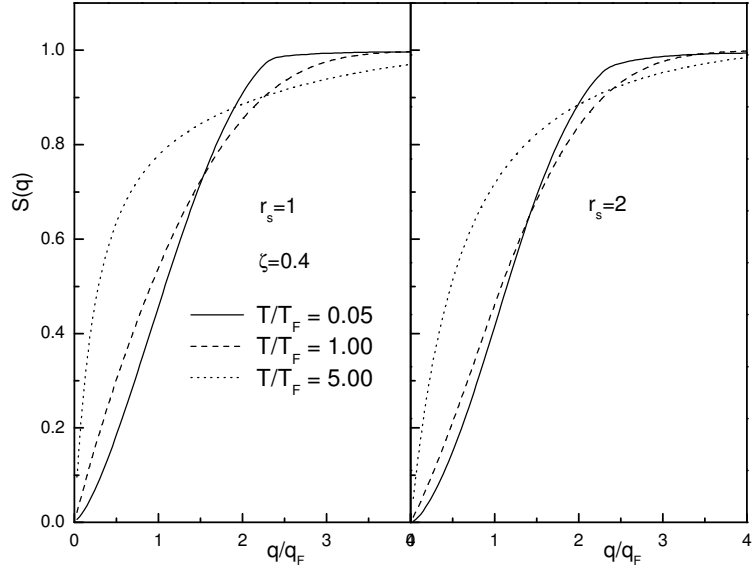

Fig. 4. Spin-averaged structure factor $S(q)$ versus $q/q_F$ for $r_s = 1$ ( left) and $r_s = 2$ ( right ) and various $T$ at $\zeta = 0.4$ .

*3.2. Pair correlation function*

The pair correlation function of a 2DEG can be calculated from the SSF using the relation (6) and the results are displayed in Figs. 5 and 6. The behaviours of the Pauli and Coulomb holes are directly seen in the PDF $g_{\sigma\sigma'}(r)$. We observe that the Coulomb hole shows weak spin-polarization dependence but appreciable dependence on the density parameter $r_s$ as in the zero-temperature case given in DT's work. From Fig. 6 we can conclude that the temperature effects on the PDF are considerable for all spin-polarizations.

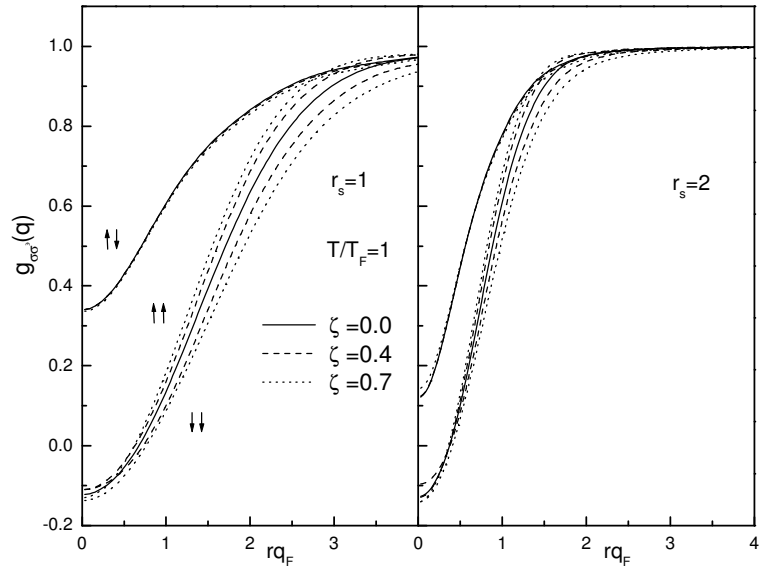

Fig. 5. Pair distribution functions $g_{\sigma\sigma'}(r)$ for $r_s = 1$ ( left) and $r_s = 2$ ( right ) and various $\zeta$ at $T = T_F$ .

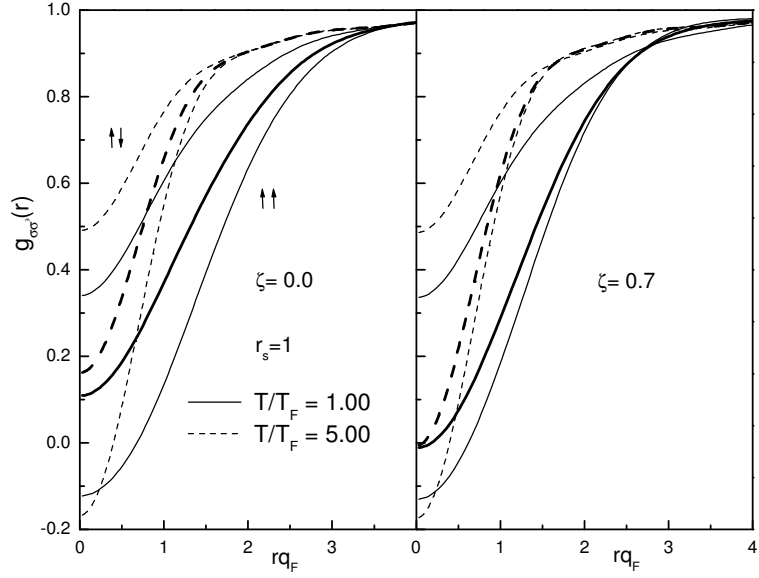

Fig. 6. Pair distribution functions $g_{\sigma\sigma'}(r)$ for $\zeta = 0$ (left) and $\zeta = 0.7$ and various $T$ at $r_s = 1$. The bold lines denote the spin-averaged PDFs $g(r)$.

*3.3. Local-field factor*

Local field factors (LFFs) $G(q)$ of unpolarized ($\zeta = 0$) and polarized ($\zeta = 1$) 2DEG at $r_s = 1$ for various $T$ are displayed in Fig. 7. We observe appreciable dependence of the LFCs on the temperature. We notice that in the paramagnetic case our results for the LFCs are similar to those given in SB's work.

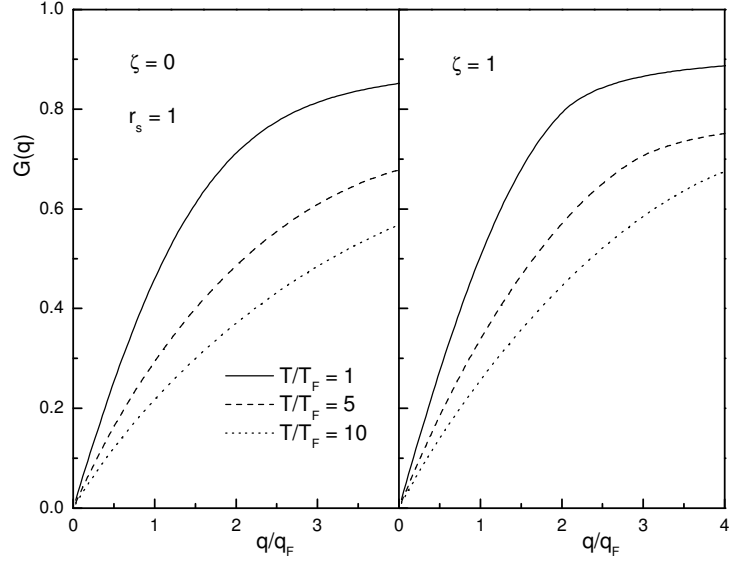

Fig. 7. Local field factors $G(q)$ versus $q/q_F$ for $\zeta = 0$ (left) and $\zeta = 1$ (right) and various $T$ at $r_s = 1$.

*3.4. Free energy*

We have calculated the exchange-correlation and total free energy of a 2DEG by using Eqs. (7) and (8). The results in the effective Rydberg ($Ryd^* = m^*e^4/2\varepsilon_o^2\hbar^2$) are given in Table 1. We note that in the paramagnetic case our results are identical to those of SB . For polarized case present results are similar to those given in DT's work for $T = 0$ and in our previous paper for finite temperatures [9] . We observe from the Table 1 that the 2DEG with temperature and density in the range studied in this paper ( $T/T_F \leq 5$, $r_s \leq 3$ ) remains in the paramagnetic phase.

Table 1
Exchange-correlation and total free energy per electron : $-F_{xc}$ and $-F$ for various $T$, $r_s$ and $\zeta$.

| $\zeta$ | 0.0 | | | 0.4 | | | 0.7 | | | 1.0 | | |
|---|---|---|---|---|---|---|---|---|---|---|---|---|
| $T/T_F$ | 0.05 | 1.0 | 5.0 | 0.05 | 1.0 | 5.0 | 0.05 | 1.0 | 5.0 | 0.05 | 1.0 | 5.0 |
| $r_s$ | | | | | | | | | | | | |
| $-F_{xc}$ | | | | | | | | | | | | |
| 1 | 1.416 | 1.267 | 0.798 | 1.465 | 1.308 | 0.831 | 1.575 | 1.421 | 0.922 | 1.723 | 1.540 | 1.000 |
| 2 | 0.759 | 0.714 | 0.492 | 0.779 | 0.732 | 0.508 | 0.823 | 0.775 | 0.544 | 0.866 | 0.801 | 0.579 |
| 3 | 0.528 | 0.507 | 0.368 | 0.541 | 0.518 | 0.380 | 0.566 | 0.541 | 0.400 | 0.583 | 0.546 | 0.418 |
| $-F$ | | | | | | | | | | | | |
| 1 | 0.424 | 2.739 | 26.387 | 0.313 | 2.523 | 25.514 | -0.093 | 2.073 | 23.553 | -0.231 | 1.180 | 19.721 |
| 2 | 0.511 | 1.082 | 6.889 | 0.491 | 1.027 | 6.679 | 0.405 | 0.937 | 6.202 | 0.377 | 0.711 | 5.259 |
| 3 | 0.418 | 0.671 | 3.211 | 0.413 | 0.653 | 3.123 | 0.376 | 0.613 | 2.915 | 0.366 | 0.331 | 2.498 |

**4. Conclusions**

In this paper using the STLS approximation we have studied the correlation characteristics of the spin polarized, finite temperature 2DEG such as the structure factor, pair correlation function, local-field factor and the Helmholtz free energy. We have shown that at low temperatures our results reduce to those of DT and in the case of finite temperature, unpolarized 2DEG our results match closely with those obtained by SB. Our results indicate that the correlation characteristics of 2DEG depend remarkably on the spin-polarization, temperature and density. We find that within the STLS approximation the 2DEG with $T/T_F \leq 5$ and $r_s \leq 3$ remains in the paramagnetic phase. The results of other properties such as the plasmon dispersion and spin susceptibility will be reported in a future publication.

*Acknowledgement*  We gratefully acknowledge the financial support from the National Program for Basic Research of Ministry of Science and Technology .